\documentclass[%
 aip,
 amsmath,amssymb,
preprint,%
]{revtex4-1}

\usepackage{graphicx}% Include figure files
\usepackage{dcolumn}% Align table columns on decimal point
\usepackage{bm}% bold math
\usepackage{xfrac}% bold math
\usepackage{float}
\newcommand{\mean}[1]{\ensuremath{\langle #1 \rangle}}

\begin{document}

\preprint{AIP/123-QED}

\title[Single-Molecule-Sensitive FRET in Droplets]{Single-Molecule-Sensitive FRET in Freely-Diffusing Attoliter Droplets}

\author{Sheema Rahmanseresht}%--
  \affiliation{Department of Physics, University of Massachusetts, Amherst, MA 01003 }%
\author{Peker Milas}%f
  \affiliation{Current address: Department of Neuroscience, University of Wisconsin, Madison, WI 53705}%
\author{Kieran P. Ramos}%
  \affiliation{Department of Physics, University of Massachusetts, Amherst, MA 01003 }
\author{Ben D. Gamari}
  \affiliation{Department of Physics, University of Massachusetts, Amherst, MA 01003 }
\author{Lori S. Goldner}
  \email{lgoldner@physics.umass.edu}
  \affiliation{Department of Physics, University of Massachusetts, Amherst, MA 01003 }

\date{\today}

\begin{abstract}
Fluorescence resonance energy transfer (FRET) from individual, dye-labeled RNA molecules confined in freely-diffusing attoliter-volume aqueous droplets is carefully compared to FRET from unconfined RNA in solution. The use of freely-diffusing droplets is a remarkably simple and high-throughput technique that facilitates a substantial increase in signal-to-noise for single-molecular-pair FRET  measurements.  We show that there can be dramatic differences between FRET in solution and in droplets, which we attribute primarily to an altered pH in the confining environment.  We also demonstrate that a sufficient concentration of a non-ionic surfactant mitigates this effect and restores FRET to its neutral-pH solution value. At low surfactant levels, even accounting for pH, we observe differences between the distribution of FRET values in solution and in droplets which remain unexplained. Our results will facilitate the use of nanoemulsion droplets as attoliter volume reactors for use in biophysical and biochemical assays, and also in applications such as protein crystallization or nanoparticle synthesis, where careful attention to the pH of the confined phase is required.  
\end{abstract}

\maketitle

Single-molecular-pair fluorescence resonance energy transfer (spFRET) is widely used in molecular biophysics to understand folding, binding, and structural changes in proteins\cite{Nettels2012} and RNA.\cite{Li2014}  In the simplest and most frequently used application of spFRET, fluorescent photons are detected as a molecule diffuses through a femtoliter-volume confocal detection region. The number of photons detected depends on the brightness of the molecule and length of time spent (dwell time) in the detection volume, typically  $< 1$ ms. To some extent, the dwell time, and therefore the signal, can be increased by adding sucrose or glycerol to increase the viscosity of the solution. An increase in the detection volume also increases the dwell time, but the gain in signal is offset by a concurrent increase in background that limits this option.

Alternatively, the dwell time of a biomolecule can be dramatically increased, and the background minimized, by confinement in a nanocontainer that is larger than the biomolecule but smaller than the detection volume. Two common candidates for molecular confinement are droplets\cite{Reiner2006} and liposomes;\cite{Boukobza2001} their relative merits have been discussed elsewhere.\cite{Tang2009}  It has long been assumed that confinement does not perturb FRET measurements; here we test that assumption for molecules confined in water-in-perfluorinated-liquid nanodroplets.

The Stokes-Einstein diffusivity for a spherical particle is $D=k_B T/6\pi \eta r$, where $r$ is the hydrodynamic radius of the particle, $k_B$ is the Boltzmann constant, $T$ is the temperature and $\eta$ is the dynamic viscosity.  The dwell time $\tau \propto w^2/D$ where $w$ is the waist of the confocal detection volume (260 nm), so that $\tau$ scales with the radius of the particle and the viscosity of the medium.  The droplets used in this study had a log-normal size distribution, with $\mean{r}=101$~nm to $135$ nm as measured both by absorption (Mie scattering) and dynamic light scattering (DLS). For a typical sample, $\mean{r}=118\pm 10$~nm with 95\% of droplets between $r=76$~nm and $r=175$~nm.   Corresponding volumes were 1.8 aL - 22 aL with a mean of 6.9~aL, resulting in $\tau\approx 12$~ms for droplets in FC-77 (Flourinert, 3M) and $\tau\approx 38$~ms for droplets in FC-40 (Flourinert, 3M); this should be compared with a $\tau\approx 375$~$\mu$s for a molecule with $\mean{r}=5$~nm in water.   

At a nominal concentration $\le 20$~nM, 6.9~aL droplets contain $\approx 0.08$ molecules on average, and the probability of finding more than 1 molecule in a droplet is $\le 0.003$.  Empty droplets are not detected. Biomolecules confined in these droplets have been observed to rotate freely\cite{Tang2008, Jofre2007} and show no evidence of sticking at the perfluorinated boundaries (Fig.~S4).\cite{2015APL_SI} This observation does not rule out more subtle interactions with the boundary.

The confined molecule is 16-base-pair duplex RNA (IDT), with  Cy3 and Cy5 (Glen Research) at the 5$^\prime$ termini,\cite{Milas2013} prepared in 20 mM Tris with 200~mM NaCl. For measurements on RNA unconfined in solution, the buffer contained  50 or 100~pM dsRNA with 100~nM protocatechuate-3,4-dioxygenase (PCD) and 2~mM protocatechuic acid (PCA)  for oxygen getting\cite{Aitken2008} and 1~mM methylviologen (MV) for triplet quenching.  For use in droplets the buffer was prepared at pH 7.8, and contained dsRNA at 10 or 20~nM with 10~mM PCA, 50~nM PCD and 1~mM MV.  Droplets were formed by adding  2~$\mu$L of this RNA-containing buffer solution to 200~$\mu$L of a continuous phase consisting of degassed Fluorinert\cite{2015APL_SI} with a triblock copolymer surfactant [perfluoro polyether (PFPE)--polyethylene glycol--PFPE, RainDance]\cite{Holtze2008} at a concentration of 0.1\%-2\% (w/w). After shaking, the mixture was placed in an ultrasonic bath  (Branson 1510) for 2-4 minutes, forming an emulsion. For all measurements, $\approx 50~\mu$L of sample was withdrawn and placed between a coverslip and microscope slide separated by double-sided sticky tape, which was sealed with grease or wax.  Data were acquired on an Olympus IX50 microscope modified for single-molecule confocal detection with a 50~$\mu$m pinhole.  A UPlanSApo $60\times$, 1.2 NA water immersion objective was used for fluorescence excitation and collection of emitted photons. The donor dye (Cy3) was excited at 514~nm (Argon-Krypton laser) with a nominal power at the scope entrance of  50~$\mathrm{\mu W}$. Fluorescent photons were split into two channels (donor, acceptor) and detected using single-photon-counting avalanche photodiodes ($\tau$-SPAD by Picoquant). Photon timing information was recorded with 8 ns resolution.\cite{Gamari2013} 

\begin{figure}[h]
  \centering
  \includegraphics[width = 3in]{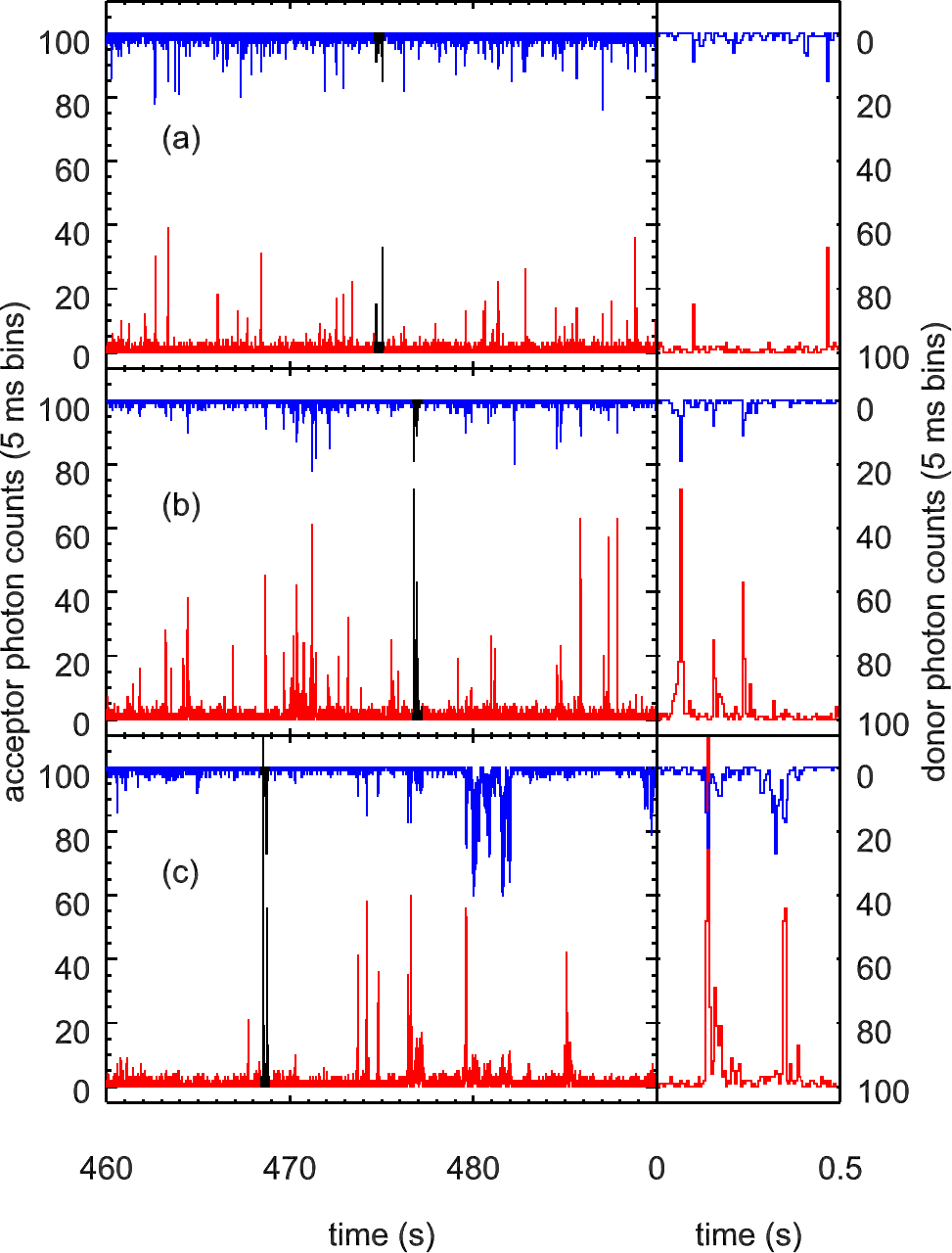}
\caption{A comparison of fluorescence from doubly-labeled RNA duplexes (a) in solution (b) confined in droplets diffusing in FC-77 and (c) confined in droplets diffusing in FC-40. The acceptor channel is plotted in red, and the donor-channel is plotted upside-down in blue, with the associated axis label on the right. A 30 s portion of the 25-75 min long data sets are shown on the left; the small panels on the right are an 0.5 s expansion of the data colored black in the left panel.}
  \label{fig:raw}
\end{figure}

A comparison of fluorescence from RNA unconfined in solution to RNA confined in droplets is shown in Fig.~\ref{fig:raw}. Photons in the donor and acceptor channels are binned in 5~ms intervals and plotted in blue and red, respectively.   The smaller panels on the right are 0.5~s expansions of the data colored black in the left panel. The peaks correspond to molecules diffusing across the detection volume. As expected, $\tau \ll 5$~ms for RNA in solution, Fig.~\ref{fig:raw}(a), so the peaks typically consist of only one or two above-background bins.  For molecules confined in aqueous droplets in FC-77, Fig.~\ref{fig:raw}(b), or FC-40, Fig.~\ref{fig:raw}(c), $\tau$ is substantially larger, with many more photons detected.  Fluorescence correlation spectroscopy was used to confirm that the dwell times were consistent with the predictions of Stokes-Einstein.

In FRET, an excited donor dye transfers its energy to a redder acceptor dye with an efficiency given by $E =[1 + \left(\sfrac{R}{R_{F}} \right)^6]^{-1} $, where $R$ is the distance between dyes and the F\"orster radius $R_F\approx 5.8$ nm for this system at neutral pH.\cite{Milas2013} Following convention,\cite{Gopich2012a} we report the histogram of the closely related proximity ratio:
\begin{equation}
P = {\frac{N_A}{N_A+N_D}}
\label{eq:P}
\end{equation}
where $N_A$ and $N_D$ are the number of photons in the acceptor and donor channels, respectively, in a given bin.  $P$ differs from $E$ only due to background, crosstalk, and differences in quantum yield or collection efficiency of the two dyes.\cite{Gopich2012a} Defining $N_t=N_A+N_D$, proximity histograms are formed using all bins with $N_t$ greater than a threshold, $N_{th}$, given below.

Proximity ratio histograms for a bin time of 2 ms are shown in Fig.~\ref{fig:prox0}.  Histograms were fit with the sum of three beta probability distribution functions (PDFs). The donor-only population, with $\mean{P} \approx 0.15$ due to crosstalk, was removed by sampling.\cite{2015APL_SI}  The best fit to the FRET peaks, and the component beta PDFs, are plotted with black and grey lines respectively. Fit parameters and peak statistics are given in Table~I. 

The proximity histogram for RNA unconfined in solution at pH 7.8 with $N_{th}=25$ is shown in Fig.~\ref{fig:prox0}(a). Data taken at pH 7.8 were indistinguishable from pH 7.0.  It is not possible to substantially increase the threshold from $N_{th}=25$; only 138 bins have $N_{t}\ge 50$.

\begin{figure}[h]
  \centering
 \includegraphics[width = 3in]{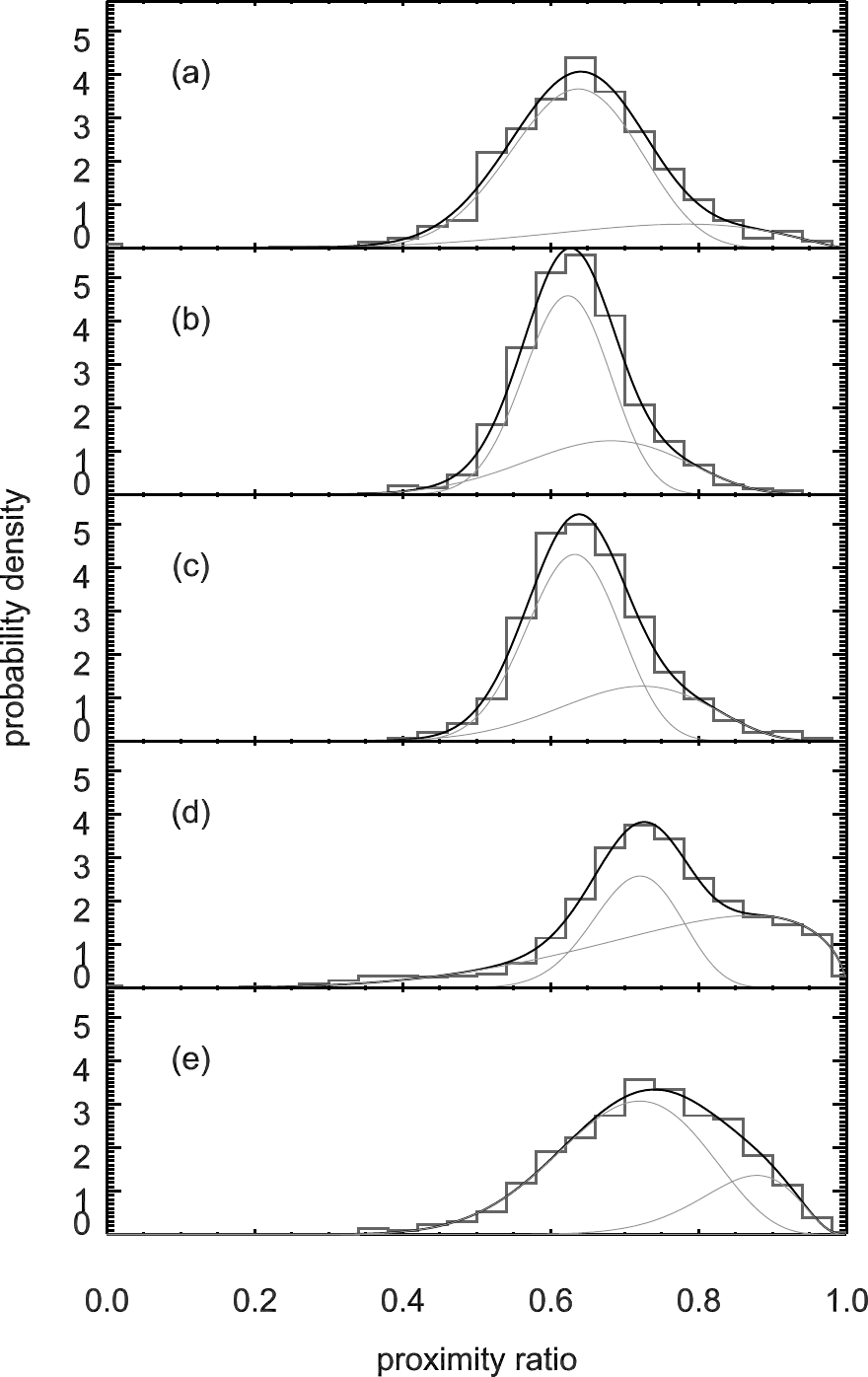}
\caption{Proximity histograms and fits for RNA (a) in solution at pH 7.8 with $N_{th} =25$; (b) in droplets in FC-40 with 2\% (w/w) surfactant and $N_{th}=75$; (c) in droplets in FC-40 with 1\% (w/w) surfactant and $N_{th}=75$; (d) in droplets in FC-40 with 0.1\% (w/w) surfactant and $N_{th} =50$; (e) in solution at pH 5 with $N_{th}=60$. See Table~I for fit parameters.}
  \label{fig:prox0}
\end{figure}

Figs.~\ref{fig:prox0}(b,c,d) are the proximity histograms for RNA in droplets in FC-40 with 2\%, 1\%, and 0.1\% (w/w) surfactant, respectively.  With 2\% and 1\% surfactant, the main peak in the droplet histogram has $\mean{P}$ that is indistinguishable from that of the pH 7.8 solution data within the $\pm 0.01$ uncertainty attributable to drift in the optics.  However, the width of the histograms, which determines statistical uncertainty and the ability to discern heterogeneous populations, is substantially narrower for droplet data. This is a result of the increase in signal afforded by droplet confinement, which allowed a threefold increase in $N_{th}$ over that of the solution data (Table~I).  A homogeneous peak in the proximity histogram ({\it i.e.}, from a Poisson emitter) has a shot-noise limited variance\cite{Gopich2012a} ${\sigma_s}^2 = [{\mean{P}\left( 1-\mean{P}\right )}]/{\mean{N_t}}$.  The factor of 2.6 increase in $\mean{N_t}$ gave a 60\% decrease in ${\sigma_s}$.

In all cases, the standard deviation of the distributions, $\sigma_d$, is larger than $\sigma_s$ (Table~I). In general we attribute this, and the need for more than one beta PDF, to photophysical transitions which are known to heterogeneously broaden proximity histograms of cyanine dyes.\cite{Kalinin2010}  

All previous reports of spFRET from droplet-confined molecules\cite{Reiner2006, Hicks2010, Goldner2010} used optically trapped droplets and the water-soluble surfactant Triton X-100. In these conditions,\cite{Goldner2010} and at 0.1\% fluorinated surfactant, FRET in droplets, Fig.~\ref{fig:prox0}(d), differs dramatically from solution FRET near neutral pH.  The histogram is both shifted from, and more distinctly heterogeneous than, that of the pH 7.8 solution data.  For the peak at lower $P$, repetitions of the low-surfactant droplet measurements ({\it e.g.},~Fig.~S5)\cite{2015APL_SI} gave the same $\mean{P}$ within the $\pm 0.01$ uncertainty. The second population had larger $\mean{P}$ and smaller $\sigma_d$ for FC-77 than FC-40,  Table S II.\cite{2015APL_SI} The relative amplitude of the two peaks varied substantially; in rare cases the higher FRET peak was the larger of the two (Fig.~S5).\cite{2015APL_SI} This RNA molecule has only one structure and photophysical effects should not cause such an obvious heterogeneity.

\begin{table}
\begin{tabular}{|ll|llllllll|}
\hline
 $\mean{N_{th}}$ & Fig & $A$ & $\alpha$  &  $\beta$  &  $\mean{P}$ & $\sigma_d$ & $ \sigma_s$ & bins & $\mean{N_t}$  \\
\hline
  25 	&	 1a 	&	   0.80 	&	   19.63 	&	   11.61 	&	  0.628 	&	  0.094 	&	  0.083 	&	   1169 	&	    37.0 \\
  25 	&	 1a 	&	   0.20 	&	    6.32 	&	    2.49 	&	  0.718 	&	  0.149 	&	  0.078 	&	    303 	&	    37.7 \\
  75 	&	 1b 	&	   0.67 	&	   43.39 	&	   26.64 	&	  0.620 	&	  0.057 	&	  0.050 	&	    898 	&	    99.6 \\
  75 	&	 1b 	&	   0.33 	&	   13.04 	&	    6.64 	&	  0.663 	&	  0.106 	&	  0.048 	&	    472 	&	   100.5 \\
  75 	&	 1c 	&	   0.68 	&	   37.28 	&	   22.10 	&	  0.628 	&	  0.061 	&	  0.048 	&	    499 	&	   104.5 \\
  75 	&	 1c 	&	   0.32 	&	   14.20 	&	    6.05 	&	  0.701 	&	  0.108 	&	  0.046 	&	    241 	&	   104.5 \\
  50 	&	 1d 	&	   0.39 	&	   40.65 	&	   16.39 	&	  0.713 	&	  0.067 	&	  0.052 	&	   2055 	&	    85.1 \\
  50 	&	 1d 	&	   0.61 	&	    5.00 	&	    1.59 	&	  0.759 	&	  0.162 	&	  0.049 	&	   3369 	&	    87.0 \\
  60 	&	 1e 	&	   0.78 	&	   14.24 	&	    6.15 	&	  0.699 	&	  0.102 	&	  0.052 	&	    449 	&	    83.3 \\
  60 	&	 1e 	&	   0.22 	&	   22.67 	&	    4.00 	&	  0.850 	&	  0.067 	&	  0.040 	&	    126 	&	    83.8 \\
\hline
\end{tabular}
\caption{Statistics and best fit parameters for the proximity histograms of Fig.~2. $N_{th}$ is the threshold number of photons per bin, Fig is the corresponding figure, $A$ is the amplitude of the beta PDF, $\mu$ and $\beta$ are the beta PDF parameters defined in the supplement,\cite{2015APL_SI} $\mean{P}$ is the mean proximity ratio, $\sigma_d$ and $\sigma_s$ are the actual and shot-noise limit of the peak standard deviations, bins are the number of bins under the peak, and $\mean{N_t}$ is the mean number of photons per bin for that peak.}
\label{table:I}
\end{table}

In searching for an explanation for the FRET shift of Fig.~\ref{fig:prox0}(d), we discovered that FRET for this system is pH dependent.   Fig.~\ref{fig:prox0}(e) is a proximity histogram for RNA in solution at pH 5 that shows a similar shift in FRET; pH 4, 5, and 6 FRET data were indistinguishable (Fig.~S6).\cite{2015APL_SI} As discussed below, we also discovered that that low-surfactant droplets are acidic; we believe this explains the observed shift of the proximity ratio in droplets.

However, with values of $N_{th}$ chosen to give similar values of $\mean{N_t}$ Figs.~\ref{fig:prox0}(d and e), we see immediately that the two proximity histograms still differ, with a more obvious splitting of the peak in the droplet data. Hypothesizing that the droplet interface might play a role in this difference, we looked for a correlation between proximity ratio and droplet size using photon-burst time length\cite{2015APL_SI} as a proxy for size.  Pearson's coefficients for burst time and $\mean{P}$ in a burst were between $-0.1$ and $+0.1$, indicating no correlation. However, the use of burst length as a proxy for size is imperfect; we cannot rule out that surface effects play a more subtle role.

We propose that changes in the F\"orster radius $ {R_F}^6 = {9 c^4 J \eta_D \kappa^2}/{8 \pi n^4}$, resulting from a modified pH in low-surfactant droplets, can account for the observed shift in $\mean{P}$. Here $n$ is the solvent's refractive index, $c$ is the speed of light, $\eta_D$ is the quantum yield of the donor dye in the absence of the acceptor,  $\kappa$ is the orientation term in a dipole-dipole interaction, and $J$ is a spectral overlap intergral.  A spectral shift large enough to substantially change $J$ would likely degrade the fluorescence signal into either the donor or the acceptor channel: instead, the brightness of droplet-confined dyes increases in both channels at low surfactant concentration, Fig.~\ref{fig:PCH}. The index difference between FC-40 (FC-77) and water is small, only 0.04 (0.05), and so it cannot significantly affect $R_F$ or dye lifetime.  This leaves changes in $\eta_D$ and/or $\kappa$ as potential causes of the shift in FRET. Both of these parameters are sensitive to changes of the dye conformation on RNA, which plausibly depends on pH.  

Cyanine dyes are insensitive to changes in pH between 4 and 10.\cite{Mujumdar1993}  However, the phosphates along the RNA backbone, which carry a double negative charge at pH 7, become singly ionized near pH 6. Also, while most ribonucleotides have a $\mathrm{pK_a} < 4$,  cytidine monophosphate has a $\mathrm{pK_a}$ of 4.5; the phosphate backbone further increases the $\mathrm{pK_a}$.\cite{Bloomfield2000}   The RNA used here was studied using MD simulations, which showed that the dyes are primarily base-stacked on the ends of the duplex at neutral pH.\cite{Milas2013}  The RNA has two C-G pairs at each end, and it is possible that protonation occuring at lower pH causes fraying of the RNA or otherwise affects the stacking of the dyes on the RNA.  Should fraying occur, cyanine dyes can intercalate into single strands, becoming substantially brighter (larger $\eta_D$).\cite{Randolph1997}   It seems likely that the shift in FRET at low pH occurs due to a modified interaction between the dyes and the RNA that causes either a change in the dye brightness (which changes $R_F$) and/or a change in the interdye distance and orientation. 

The proposed change in brightness was indeed observed using photon-counting histograms\cite{Chen1999, Huang2004} (PCHs) of donor-only (Cy3) labeled RNA. The PCHs shown in Fig.~\ref{fig:PCH} were taken on the same day under identical conditions with an excitation power of 50~$\mu$W.\cite{2015APL_SI} For RNA in solution at pH 7, Fig.~\ref{fig:PCH}(a), the data are fit well by two species, presumably Cy3 isomers but possibly different conformations of Cy3 on the RNA, one with roughly twice the brightness and $< 10$\% the population of the other.  Similar results were obtained at pH 7.8. Below pH 7 a new, brighter species emerges: three populations are required for a good fit. In solution at pH 4, Fig.~\ref{fig:PCH}(b), this new species is roughly eight times brighter than the dimmest species and comprises roughly 4\% of the population.  For droplets with 0.1\% (w/w) surfactant, the situation is similar, Fig.~\ref{fig:PCH}(c) and (d); the new species is 5 to 6 times brighter than the dimmest species, and comprises at most 5.5\% of the population.  Differences between FC-77 and FC-40 are mostly insignificant; a complete set of PCH fitting parameters is given in Table SIII.\cite{2015APL_SI} 

\begin{figure}[h]
  \centering
 \includegraphics{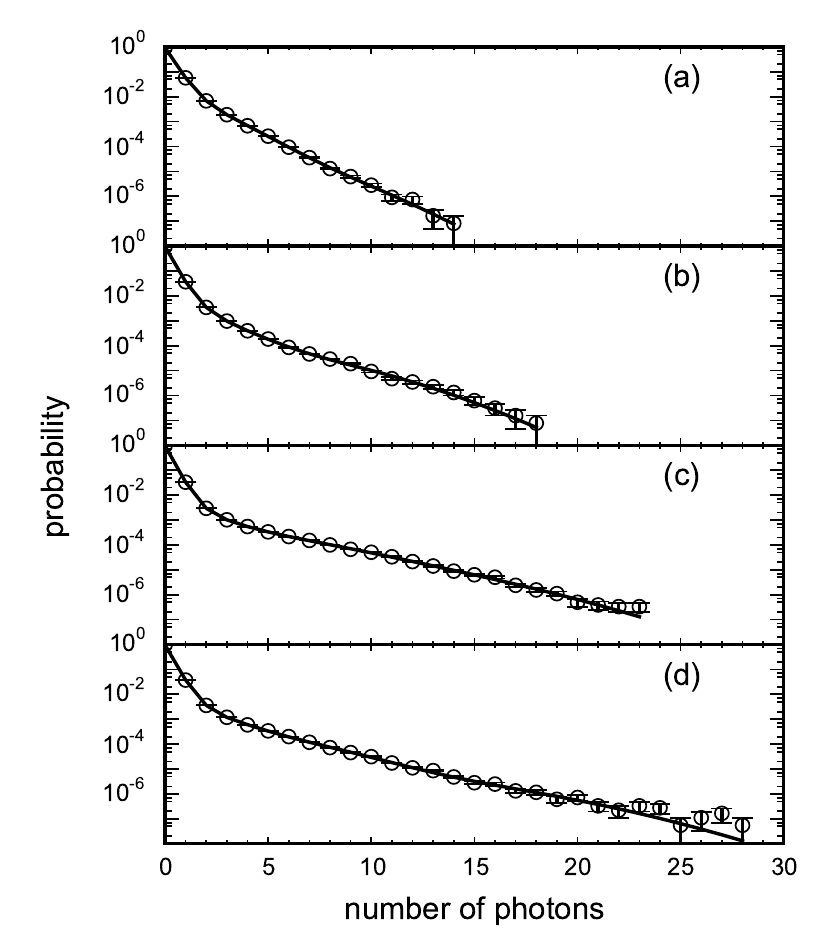}
\caption{PCH for donor-only-labeled RNA (a) in solution at pH 7 with a two species fit (line) with $\chi^2=0.7$ ; (b) in solution at pH 4 with a three species fit (line) with $\chi^2=1.3$; (c) in droplets in FC-77 and 0.1\% surfactant with a three species fit (line) with $\chi^2=1.1$; and (d) in droplets in FC-40 and 0.1\% surfactant with a three species fit (line) with $\chi^2=1.4$. $ \chi^2$ is calculated per degree of freedom and the bin time is  200 $\mu s$.}
 \label{fig:PCH}
\end{figure}

Using pH sensing dyes, we confirmed that a non-ionic surfactant affects the pH in droplets, contrary to expectations. Starting with buffer at pH 7.8, droplets with 0.1\% (w/w) surfactant had a confined-phase pH of 5.5, while above 1\% surfactant the pH was near 7.3.  Details will be reported elsewhere. The acidity of small water droplets with little or no surfactant was unexpected but should not be surprising. The zeta potential of particles, oil droplets, and air bubbles in pure water is known to be negative, a phenomena widely ascribed to the autolysis of water and sequestration of hydroxide ions near or on the water boundary.\cite{Beattie2009a} It has been shown that for oil-in-water emulsions, as the interfacial area increase ({\it e.g.} by decreasing droplet size), the pH of the aqueous phase decreases.\cite{Beattie2004}  To maintain the pH of the continuous phase, it is necessary to titrate in enough NaOH to provide one OH- for every 3 nm$^2$ of surface. For a 100 nm diameter droplet, this corresponds to about $4\times 10^4$ hydroxides sequestered at or very near the surface, more than enough to account for the observed change in the pH of the confined phase. 

In conclusion, FRET in droplets with $\ge 1\%$ surfactant offers better signal to noise and is otherwise the same as FRET from molecules unconfined in solution.  With $0.1\%$ surfactant, droplets becomes acidic and $\mean{P}$ exhibits a shift that is similar in droplets and in bulk solution at low pH. 

However, FRET from RNA confined in low surfactant droplets has greater heterogeneity than is observed in solution at low pH. One plausible hypothesis is that the kinetics are different. If the two populations seen in the proximity histograms represent the same dye isomers or dye conformations on RNA, then the more distinct splitting in Fig.~\ref{fig:prox0}(d) might indicate that kinetic transitions are slower in droplets than in bulk solution. When transitions are fast compared to the bin time, heterogeneities are washed out. Further investigation of kinetics in droplets would be needed to test this hypothesis. It seems likely that interactions at the water boundary play a role. Even if a droplet is at neutral pH, there will be a space charge layer near the surface.  This might affect the kinetics, as well as the rotational and translation diffusion of the molecule. 

The authors thank John Randolph at Glen Research, Brian Hutchison at RainDance Technologies, Anthony Dinsmore, Adrian Parsegian, and Rudi Podgornik at UMass for useful and illuminating discussion.  This work was funded by NSF MCB-0920139 and NSF DBI-1152386.

\bibliography{Goldner_refs_August_2014}

\newcommand{\lburst}{\ensuremath{\lambda_\mathrm{Burst}}}
\newcommand{\lbg}{\ensuremath{\lambda_\mathrm{BG}}}
\newcommand{\wburst}{\ensuremath{w_\mathrm{Burst}}}
\newcommand{\wbg}{\ensuremath{w_\mathrm{BG}}}
\renewcommand{\figurename}{Fig.~S}
\renewcommand{\tablename}{Table.~S}

% Hack for making SOM Equations Conform to Science Format
%
% e.g. (S1), (S2), etc
% Requires AMS
\makeatletter %% With ams
\def\tagform@#1{\maketag@@@{(S\ignorespaces#1\unskip\@@italiccorr)}}
\makeatother

%******************************************************************************************************

\pagebreak

\uppercase{\textbf {\center Supplementary Information for Single-Molecule-Sensitive FRET in Freely-Diffusing Attoliter Droplets}}

\maketitle

\section{Confocal Image of a Large Droplet}
Consistent with our earlier work using Green Fluorescent Protein\cite{Tang2008} and DNA,\cite{Jofre2007} we find no evidence that RNA sticks at the droplet boundaries.  Confocal scanning images of large (micron) droplets provide evidence that the RNA fills the entire volume with no obvious preference for the interface,  Fig.~S4. In this figure, the 16 base-pair duplex RNA labeled with Cy3 at a $5'$ termini (identical to that used for donor-only PCH measurements in the text) was prepared at 16.7 $\mu$M in 20~mM Tris buffer with 200~mM NaCl.  Droplets were created by adding 2 $\mu$L of RNA sample into 200 $\mu$L perfluorinated oil and surfactant solution as described below and in the text. Sample was then shaken for 1-2 minutes, resulting in much larger droplets suitable for investigation by confocal scanning. Droplets were imaged at or very near to a glass boundary; the confocal image and corresponding line plot are centered at least one micron above a coverslip.

\begin{figure}[h]
  \centering
  \includegraphics[width = 3in]{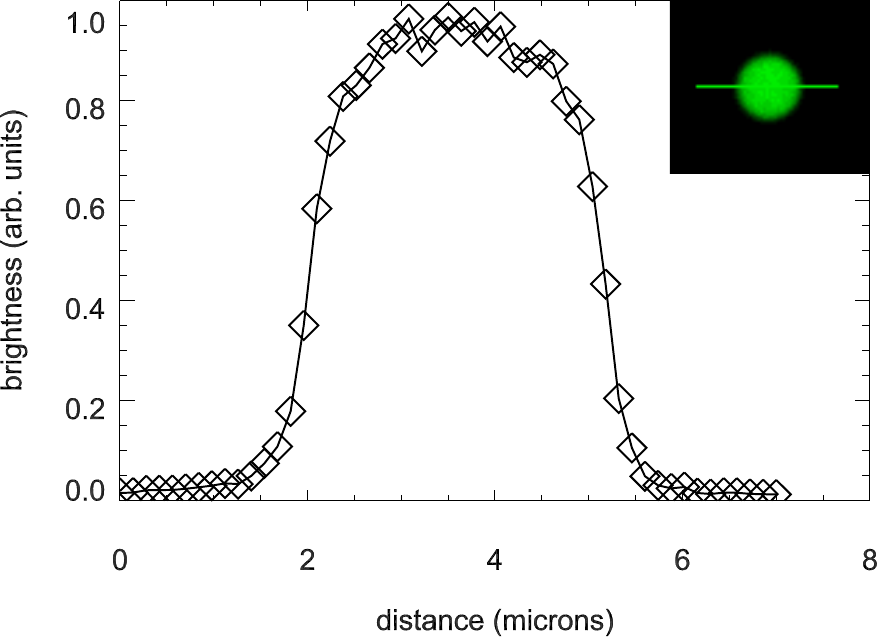}
\caption{Confocal scanning image of very large droplet in FC-40, as described above. }
  \label{fig:S4}
\end{figure}

\section{Methods}

\subsection{Sample Preparation} The RNA 16 base oligo 5$^\prime$-Cy3-C-G-A-G-U-G-A-C-C-A-G-U-G-A-G-C-3$^\prime$ and its complement with and without a Cy5 at the 5$^\prime$ terminus, were obtained from IDT. Cy3 and Cy5 are indocarbocyanine dyes supplied by Glen Research.  Donor (Cy3) and acceptor (Cy5) labeled ribonucleotides were prepared in 20 mM Tris at pH 7.8 with 200~mM NaCl. In this buffer, RNA at 0.75~$\mu$M was heated to 90$^\circ$~C in 5 minutes and then incubated at 90$^\circ$~C for 60 minutes before cooling to 4$^\circ$~C over 60 minutes.  For measurements on freely diffusing molecules, 100~nM protocatechuate-3,4-dioxygenase (PCD) and 2~mM protocatechuic acid (PCA) were mixed in 20~mM Tris with 200 mM NaCl and incubated for 10 minutes.  PCA/PCD functions as an enzymatic oxygen quenching system.\cite{Aitken2008} The dsRNA samples were diluted in this buffer to a concentration of 50~pM or 100~pM  with 1~mM methylviologen (MV). 

\subsection{Droplet Preparation} The dsRNA sample was prepared in emulsion  as follows: 2~$\mu$L of dsRNA at 10~nM or 20~nM with 10~mM PCA, 50~nM PCD and 1~mM MV  was added to a 200~$\mu$L of a continuous phase consisting of degassed perfluorinated oil (3M Fluorinert FC-40 or FC-77) with $10^{-3}$ v/v perfluorinated surfactant (RainDance).\cite{Holtze2008} After shaking, the mixture was sonicated for 2-4 minutes in an ultrasonic cleaner (Branson 1510), which formed the emulsion. FC-77 is primarily 2-(nonafluorobutyl)heptafluorofuran, with average molecular mass of 416, viscosity of 1.3~cP and refractive index of 1.28. FC-40 is primarily perfluorotributylamine, with average molecular mass 650, viscosity of 4.1~cP, and  refractive index of 1.29.  Note that in both cases the refractive index is near but lower than that of water ($n=1.33$).

Approximately 50~$\mu$L of emulsion was withdrawn and placed between a coverslip and microscope slide separated by double-sided sticky tape, which was then sealed with silicone vacuum grease or valap. 

Degassing of the perfluorinated oils was achieved by the freeze-pump-thaw method. Perfluorinated oils were placed in a sealed Schlenk flask and frozen in liquid nitrogen.  The flask was then opened to vacuum and pumped to 12 mtorr, re-sealed and thawed in a warm water bath.  After sitting for 30 minutes the process was repeated up to five times. During the final thaw cycle, the flask was filled with dry Nitrogen gas at slightly positive pressure.

\subsection{FRET Measurements} 
 In fluorescence resonance energy transfer (FRET), an excited donor dye transfers its energy to a redder acceptor dye if  the molecules are sufficiently close:
\begin{equation}
   E = \frac{1}{1 + \left( \frac{R}{R_{F}} \right)^6} ,
\label{eq:E}
\end{equation}
where $E$ is the energy transfer efficiency, $R$ is the distance between dyes and the F\"orster radius
$R_F$ is given by \cite{Clegg1992,Novotny2006}
\begin{equation}
  {R_F}^6 = \frac{9 c^4 J \eta_D \kappa^2}{8 \pi n^4} .
\label{eq:RF}
\end{equation}
In this expression, $n$ is the solvent's refractive index, $c$ is the speed of light, $\eta_D$ is the quantum yield of the donor dye in the absence of the acceptor, and $\kappa$ is a factor that describes the relative orientation of the dyes;  $\mean{\kappa^2} = 2/3$ for freely rotating dyes. The symbol $J$ describes the overlap of the donor emission and acceptor absorption spectra.\cite{Novotny2006} With the assumption of fast and freely rotating dyes, and sufficiently rapid fluctuations in $R$, Eq.~S2 can be directly used to determine the distance between disparate points in a molecule; more frequently it is used to qualitatively observe global changes in molecular structure or binding.  Cyanine dyes on RNA are not freely rotating\cite{Milas2013} and so FRET cannot be calculated from Eq.~S2.\cite{Gopich2012a}  From MD simulations on this duplex\cite{Milas2013} at neutral pH we know that $\mean{\kappa^2}$ can be quite low.  The shift to higher FRET at low pH may therefore be the result of a larger value of $\mean{\kappa^2}$ for dyes on protonated RNA.

In considering the causes of the difference between FRET at low and high surfactant concentrations, we also examined potential experimental artifacts. Efforts to measure crosstalk and the parameter $\gamma$, that describes the relative quantum yields and collection efficiencies of the two fluorescence channels,\cite{Gopich2012a} gave similar results for all data. Additional surfactant sometimes raised the background slightly, but the resulting difference in $P$ is insignificant. FRET for this system was also insensitive to salt concentration from at least 100 mM to 800 mM (data not shown), so it is unlikely that any difference in the salt conditions in droplets resulting from different conditions at the surface could explain the observed shift.

\subsection{Proximity Histograms and Fit Results}
Proximity ratio histograms were modeled by the sum of up to three beta probability distribution functions (beta PDFs), representing the donor-only peak and up to two distinct FRET peaks:
\begin{equation}
P(x\vert \{A_i,{\alpha}_i, {\beta}_i\}) =\sum_{i=1}^{2\, or\, 3}A_i  \frac{x^{\alpha_i-1}(1-x)^{\beta_i-1}} {\mathrm{B}(\alpha_i,\beta_i)}.
\label{eq:proxfit}
\end{equation} The normalization constant is the beta function $\mathrm{B}(\alpha,\beta)=\Gamma(\alpha)\Gamma(\beta)/\Gamma(\alpha+\beta)$.  Best fit parameters were determined by the method of nonlinear least squares.  Each data bin was then assigned to a specific peak $i$ as follows.  From the best fit parameters, the weighted probability of each bin to be in state $i=0$ is first calculated.  A variate is then drawn from a uniform distribution on the interval of 0 to 1.  If this variate is greater than the probability of the bin to be in state $i=0$ then the bin is re-assigned to state $i=1$.  If there is a third state, another variate is drawn for each $i=1$ assignment, with a resulting re-assignment to state $i=2$ if the variate is greater than the probability of the bin to be in state $i=1$. Bins assigned to state $i=0$ belonged to the donor-only population and are not included in the proximity histograms here or in the text. 

For completeness, we include here an additional example of the proximity ratio histogram of RNA confined to droplets in FC-77, evaluated at various values of $\mean{N_{th}}$. The data in Fig.~S5 had the largest high-FRET peak observed; in most other cases, the lower $\mean{P}$ peak had the larger amplitude for  $\mean{N_{th}}=75.$ 

\begin{figure}[h]
\centering

\includegraphics[]{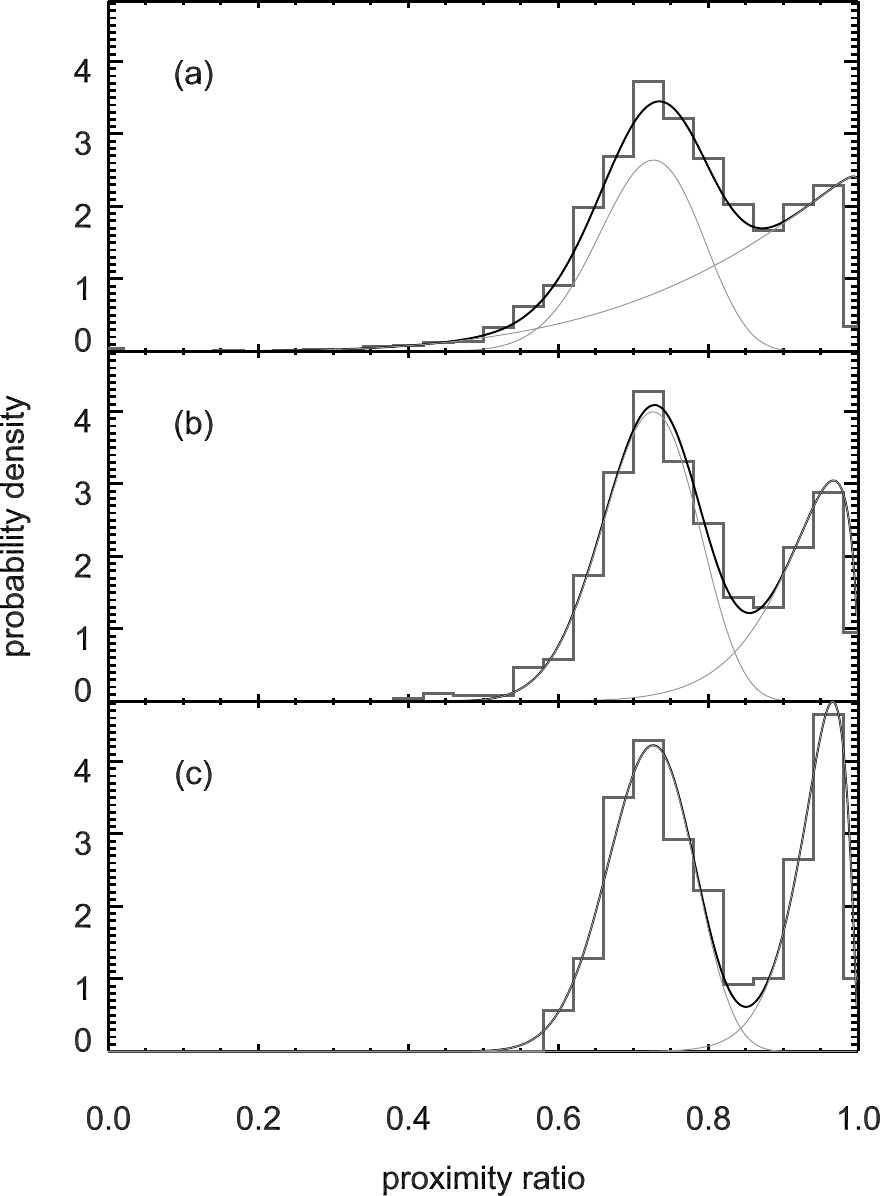}
\caption{Proximity histogram of RNA molecules confined to freely-diffusing aqueous droplets in FC-77.  Photon bin time is 2~ms. The three panels represent the same data but with different thresholds for inclusion in the histogram: (a) $N_{th} > 25$, (b) $N_{th}>50$, and (c) $N_{th}>75$.  The data are fit with beta PDFs; fit parameters are given in Table~S II.}
\label{fig:prox77}
\end{figure}

\begin{table}[h]
\centering
\begin{tabular}{|ll|llllllll|}
\hline
 $\mean{N_{th}}$ & Fig & $A$ & $\alpha$  &  $\beta$  &  $\mean{P}$ & $\sigma_d$ & $ \sigma_s$ & bins & $\mean{N_t}$  \\
\hline
  25 	&	 S2a 	&	   0.45 	&	   30.82 	&	   12.21 	&	  0.716 	&	  0.071 	&	  0.071 	&	   1478 	&	    47.1 \\
  25 	&	S2a 	&	   0.55 	&	    4.87 	&	    1.03 	&	  0.826 	&	  0.147 	&	  0.061 	&	   1676 	&	    49.5 \\
  50 	&	S2b 	&	   0.64 	&	   35.46 	&	   14.00 	&	  0.717 	&	  0.071 	&	  0.056 	&	    744 	&	    70.3 \\
  50 	&	 S2b 	&	   0.36 	&	   17.42 	&	    1.57 	&	  0.918 	&	  0.063 	&	  0.034 	&	    419 	&	    75.8 \\
  75 	&	S2c 	&	   0.61 	&	   43.62 	&	   17.07 	&	  0.719 	&	  0.055 	&	  0.047 	&	    213 	&	    92.0 \\
  75 	&	S2c 	&	   0.39 	&	   34.29 	&	    2.19 	&	  0.940 	&	  0.039 	&	  0.024 	&	    137 	&	   102.1 \\
\hline
\end{tabular}

\caption{Fit parameters for data of Fig.~S5.}
\label{tab:prox77}
\end{table}

In Fig.~S6 we show proximity histograms of RNA at three different pH in solution, on the same plot, for comparison.  Data at pH 5 was similar to pH 4 and pH 6.

\begin{figure}[H]
 \centering
  \includegraphics[width = 3in]{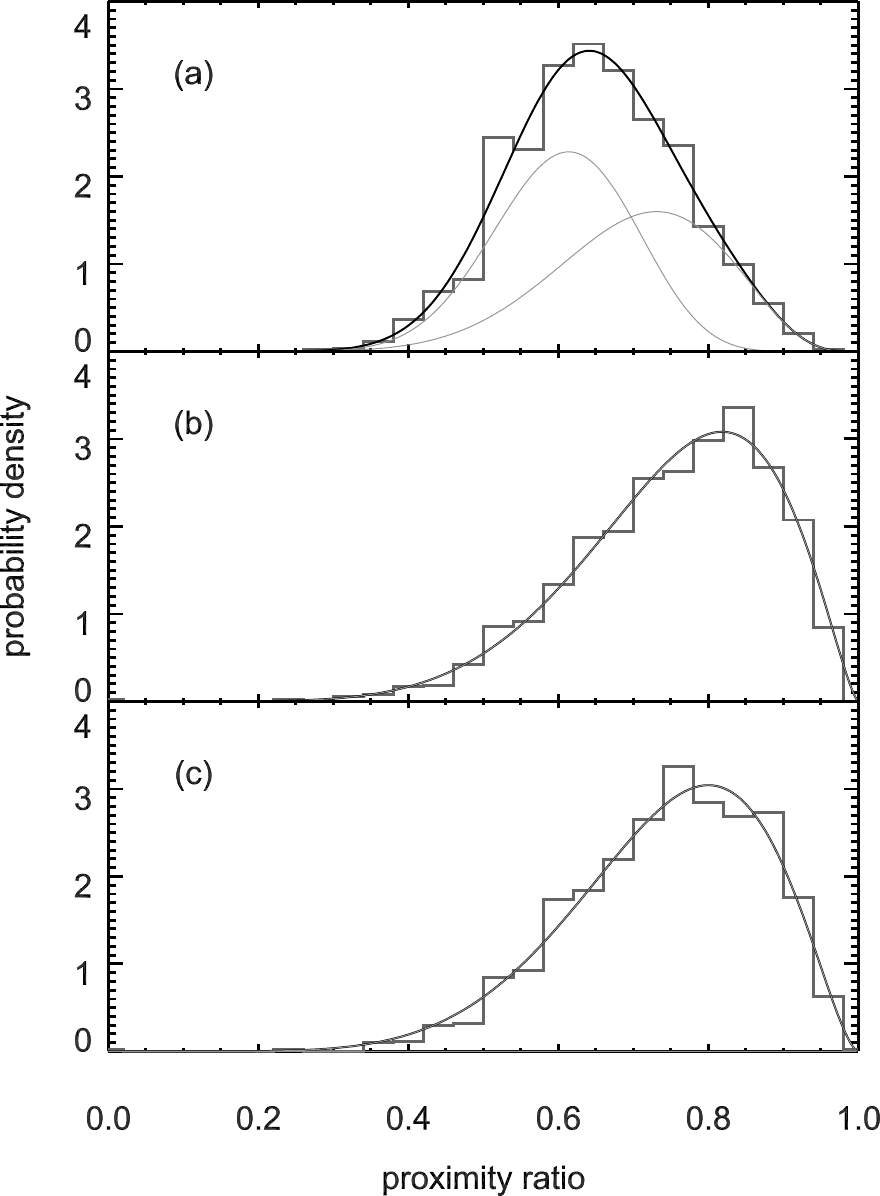}
\caption{Proximity histograms from freely diffusing RNA at (a) pH 7.0, (b) pH 6 and (c) pH 4.  Photon bin time is 2~ms, and the threshold for inclusion in the histogram is set at $N_{th} > 25$ in all three cases. The data are fit with beta PDFs, fit parameters are given in Table~S III.  Note that at this low threshold, the low pH proximity histograms are well fit by a single beta PDF.}
  \label{fig:proxsol}
\end{figure}

\begin{table}
  % \documentclass[]{nature}
% \usepackage[textsize=tiny,textwidth=70,color=blue!20,colorinlistoftodos]{todonotes}
% \usepackage{float}
% \usepackage{amsmath}
% \usepackage{braket}
% \newcommand{\mean}[1]{\ensuremath{\langle #1 \rangle}}
% \begin{document}
\begin{tabular}{|l|l|llllllll|}
\hline
 pH & threshold & $A$ & $\alpha$  &  $\beta$  &  $\mean{P}$ & $\sigma_f$ & $ \sigma_s$ & bins & $\mean{N}$  \\
\hline
 7 	&	 25 	&	  0.460 	&	   10.92 	&	    4.65 	&	  0.701 	&	  0.112 	&	 0.0855 	&	    622 	&	   29.23 \\
  	&	  	&	  0.540 	&	   16.45 	&	   10.73 	&	  0.605 	&	  0.092 	&	 0.0922 	&	    686 	&	   28.70 \\
 6 	&	 25 	&	  1.000 	&	    7.44 	&	    2.43 	&	  0.754 	&	  0.131 	&	 0.0745 	&	   2379 	&	   36.03 \\
 4	&	 25 	&	  1.000 	&	    7.58 	&	    2.64 	&	  0.741 	&	  0.131 	&	 0.0759 	&	   1272 	&	   35.71 \\
\hline
\end{tabular}
% \end{document}

\caption{Fit parameters for the data of Fig.~S6}.
\label{tab:proxsol}
\end{table}

\subsection{Photon-Counting Histogram Analysis} 
    A photon-counting histogram (PCH) is a histogram of the number of photons per bin during a photon-counting experiment.  Analysis of the PCH gives the average brightness and average number of molecules in the detection volume for multiple species. Species are distinguished only by their brightness, not by their diffusivity as in fluorescence correlation spectroscopy (FCS).  PCH is often used as a complement to FCS in the analysis of photon statistics in single-molecule-sensitive measurements.  PCHs are modeled by a super-poissonian distribution as developed by Chen {\it et al.}\cite{Chen1999} and later updated by Huang {\it et al}.\cite{Huang2004}  Here we follow the method and nomenclature of the latter.  This model assumes a cylindrically symmetric three dimensional Gaussian detection volume, with two correction parameters that describe deviations from Gaussian.  Fitting parameters therefore include the brightness $\epsilon$ and molecular concentration $\mean{n}$ for each species as well as beam-shape correction factors $F_1$ and $F_2$.  The parameter $F_1$, called the out-of-focus emission ratio, gives the ratio of the photons detected in the non-Gaussian part of the beam to the Gaussian part. When $F_1$ is large it becomes necessary to use a second parameter $F_2$ which increases the probability that a molecule in the non-gaussian part of the beam contributes two photons instead of just one.  This is the approach developed by Huang {\it et al.} \cite{Huang2004} and used to produce the fits in Fig.~3 of the text.

    For all fits the the Gaussian beam waist was taken to be 260~nm with an aspect ratio of 9:1, as suggested by a calibration of the instrument using FCS. We chose the arbitrary parameter\cite{Huang2004} $Q=6$ and used a bin time of 200 $\mu$s.

    When fitting the droplet and low pH solution data it was found that a fit with one or two species did not work well. Such fits resulted in a large $\chi^2$, and/or non-random residuals and/or very large standard errors on some of the fit parameters. A model with three species gave values of $\chi^2$ per degree of freedom near one in all cases. The pH~7 solution data were fit to two species. Fits to one species were unsatisfactory with large $\chi^2$ or residuals, and attempts to add a species gave meaningless results for the third species.  We chose to fix the parameter $F_1$ for pH~4 solution data to the same value found for the pH~7 data since this is a shape parameter that should be nominally the same for all solution data; small differences in alignment or index of refraction at low pH might account for the small change in $F_2$ that was required for a good fit. For droplet data we found it necessary to let the shape parameters vary to obtain good fits: It is reasonable to assume that the different indices of the oils and presence of the droplet may alter the shape of the detection volume.  The final values of the shape parameters were only slightly different from those found in solution.

\begin{table}
\scriptsize
\begin{tabular}{| l | c | c | c | c | c | c | c | c |}
\hline
sample  & $\mean{n_1}   $ & $\epsilon_1$ & $\mean{n_2}$ &  $\epsilon_2$ & $\mean{n_3} $ & $\epsilon_3$ &  $F_1$       &  $F_2$        \\  
\hline
 FC40    &     0.0409(5)      &   2.45(37)   &     0.0189(12)     &    9.19(55)   &     0.0035(14)     &   15.2(1.1)  &  1.27(7)     &  0.012(3)     \\
 FC77   &     0.0532(21)     &   3.32(29)   &     0.0217(21)     &    8.41(37)   &     0.0013( 3)     &   17.9(1.0)  &  1.45(4)     &  0.022(2)     \\
 pH 4    &     0.083(12)      &   1.11(69)   &     0.056(28)      &    3.24(62)   &     0.0064(13)     &    9.22(48)  &  1.38(fixed) &  0.0258(5)    \\
 pH 7    &     0.18(11)       &   2.7( 7)    &     0.0055(18)     &    6.5(5)     &         --         &      --      &  1.38(4)     &  0.048(4)     \\
 \hline
 \end{tabular}
 \caption{PCH fit parameters of the data in Fig.~3 in the text. Uncertainties are given in parentheses and represent the error on the last digits.}
 \label{table:PCH}
\end{table}

\subsection{Burst Detection}

Burst detection was accomplished using a simple Bayesian method based on photon inter-arrival times.  All the photons (both channels) are used, and the method distinguishes between photons from fluorescent bursts and photons from background.  To determine if the $i$th photon originates from a burst, the arrival times of $N$ photons on either side of the $i$th photon were examined.  Here we use a``window" with  $N=5$ photons.

Starting with the assumption of two Poisson processes, we assigned initial rates $\lburst$ and $\lbg$ associated with each.  Rather than considering directly the probability  that the $i$th photon originates from either background or burst, we consider first that fast fluctuations between the two states are unphysical; a single "burst" photon between long stretches of background photons, and the opposite, should be avoided.  We therefore consider the probability that $2N+1$ sequential photons all originate from a burst: 
\begin{equation}
P= \prod_{j=i-N}^{i+N} P(\tau_j\vert \lburst).
\end{equation}
We compare this with the probability that the same photons originate from background:
\begin{equation}
Q= \prod_{j=i-N}^{i+N} P(\tau_j\vert \lbg).
\end{equation}  
Defining
\begin{equation}
 R= \frac{\wburst P}{\wbg P+ \wburst Q},
\end{equation}
the $i$th photon is assigned to a burst if $S<R$, where $S$ is a random number uniformly distributed on the interval 0 to 1.  The weights $\wburst$ and $\wbg$ are initially set equal to 1, and after the first iteration are calculated from the sample.  This process converges by approximately 20 iterations for most data sets.

\bibliography{Goldner_refs_August_2014}

%\end{document}

\end{document}